\documentclass[%
 reprint,
 amsmath,amssymb,
 aps,
]{revtex4-1}

\usepackage{graphicx}
\usepackage{dcolumn}
\usepackage{bm}

\begin{document}

\preprint{APS/123-QED}

\title{Lorenz Gauge Condition and Its Relation to Far-Field Longitudinal Electric Wave}
\thanks{In some textbooks authors use the term ``Lorentz'' gauge condition. However, the correct name should be ``Lorenz'' gauge condition, named after Ludvig Lorenz.}%

\author{Altay Zhakatayev}
\affiliation{%
 Robotics Department, Nazarbayev University, Nur-Sultan, Kazakhstan 
}%




\date{\today}

\begin{abstract}
In this short paper, it is shown that Lorenz gauge condition leads to disappearance of long-range longitudinal electric wave emitted by an arbitrary electrical system. In any other gauge, longitudinal electric wave would be non-negligible. Implications of the obtained results are discussed. 
\end{abstract}

\maketitle


\section{\label{sec:introduction} Introduction}

Derivations of radiation of infinitesimal electric dipoles, linear traveling and standing wave antennas and other radiating systems are well known. In these derivations, it is often shown that no far-field longitudinal electric wave is emitted. However, in this paper, this commonly accepted point of view is challenged. Specifically, we show that in the general case and within the framework of classical electrodynamics, solutions yield the emission of long-range longitudinal electric wave. 

\section{Theory}
\label{sec:theory}

Our purpose in this paper is to show that Lorenz gauge condition ``destroys'' long-range longitudinal electric field or wave. For starter, we will describe briefly fundamental concepts. 

For the derivation, spherical coordinate system is used, which has unit vector $\mathbf{\hat{r}}$, $\mathbf{\hat{\theta}}$, $\mathbf{\hat{\phi}}$. $\mathbf{r}$ denotes arbitrary radius-vector with magnitude $r$, while $t$ is time. $\omega$ is angular frequency of the radiation, while $k = \frac{\omega}{c}$ is ``spatial'' frequency or wave-number. All force and potentials fields are assumed to be dependent on $\mathbf{r}$ and $t$ (e.i. $\mathbf{E} = \mathbf{E}(\mathbf{r},t)$, $\mathbf{B} = \mathbf{B}(\mathbf{r},t)$, $\mathbf{A} = \mathbf{A}(\mathbf{r},t)$, $V = V(\mathbf{r},t)$), but we will not write it explicitly in order to be concise. The term longitudinal electric field is used to denote the component of electric field $E_r$ along radial $\mathbf{\hat{r}}$ direction.

\subsection{Background}

Maxwell equations have the following form
\begin{subequations}
\begin{equation}\label{eqn:maxwell1}
    \nabla \cdot \mathbf{E} = \frac{\rho}{\epsilon_0},
\end{equation}
\begin{equation}\label{eqn:maxwell2}
    \nabla \cdot \mathbf{B} = \mathbf{0},
\end{equation}
\begin{equation}\label{eqn:maxwell3}
    \nabla \times \mathbf{E} = -\frac{\partial{\mathbf{B}}}{\partial{t}},
\end{equation}
\begin{equation}\label{eqn:maxwell4}
    \nabla \times \mathbf{B} = \mu_0 (\mathbf{J} +\frac{\partial{\mathbf{E}}}{\partial{t}}).
\end{equation}
\end{subequations}
By utilizing vector caclulus identities, the vector potential is introduced from (\ref{eqn:maxwell2}), while the scalar potential is defined by using the vector potential in (\ref{eqn:maxwell3}). As a result, the following expression is obtained for the electric field 
\begin{equation}\label{eqn:electric}
    \mathbf{E} = -\nabla V - \frac{\partial{\mathbf{A}}}{\partial{t}}.
\end{equation}
It is claimed that divergence of the vector potential can be set to arbitrary value \cite{DavidGriffiths}. Different values correspond to different gauge conditions. The frequently utilized gauge condition is called Lorenz gauge and it is defined as 
\begin{equation}\label{eqn:lorenz_gauge}
    \nabla \cdot \mathbf{A} = - \frac{1}{c^2} \frac{\partial{V}}{\partial{t}}.
\end{equation}

We will now switch to a so-called complex or $j$ notation. According to \cite{balanis2005antenna} Sec. 3.6, in the most general case the far-field terms of vector and scalar potentials have the following form
\begin{subequations}\label{eqn:general_potential}
\begin{align}
    & \mathbf{A} = (A_r(\theta,\phi) \mathbf{\hat{r}} + A_{\theta}(\theta,\phi) \mathbf{\hat{\theta}} + A_{\phi}(\theta,\phi) \mathbf{\hat{\phi}}) \frac{e^{-j(kr-\omega t)}}{r}, \\
    & V = V(\theta,\phi) \frac{e^{-j(kr-\omega t)}}{r}.
\end{align}
\end{subequations}
This can also be confirmed by checking the solution of far-field radiation of any electrodynamics system (short dipoles, long traveling wave and standing wave antennas, circular antenna and etc.). In essence, (\ref{eqn:general_potential}) means that angular ($\theta$, $\phi$) and radial ($r$) dependence of potentials can be separated or decoupled. Our focus is not on how to get the vector and scalar potentials (\ref{eqn:general_potential}), but to investigate the implications which follow from already having these potentials. 

\subsection{Derivation in Lorenz Gauge}

Let us briefly demonstrate that the derivation in \cite{balanis2005antenna} Sec. 3.6, which proves that far-field longitudinal electric wave does not exist, is valid only for one specific gauge. Similar derivations are performed in \cite{king1969antennas} Sec. 1.13 and 1.15. In the complex notation, Lorenz gauge condition becomes $\nabla \cdot \mathbf{A} = -j \omega V / c^2$, from which we can obtain the scalar potential and substitute it into (\ref{eqn:electric}). This would result in the following expression for the electric field:
\begin{equation}\label{eqn:electric2}
    \mathbf{E} = -j \frac{c^2}{\omega}\nabla (\nabla \cdot \mathbf{A}) - j \omega {\mathbf{A}}.
\end{equation}
The vector potential from (\ref{eqn:general_potential}) is utilized to find
\begin{equation}\label{eqn:identity2}
    \nabla \cdot \mathbf{A} = -A_r jk \frac{e^{-j(kr-\omega t)}}{r},
\end{equation}
where only the term with first power in $r$ in the denominator is selected. This is done because we are analyzing far-field approximation and so terms decreasing faster then $1/r$ can be neglected.
Next let us find
\begin{equation}\label{eqn:equation1}
    \nabla (\nabla \cdot \mathbf{A}) = -A_r k^2 \frac{e^{-j(kr-\omega t)}}{r} \mathbf{\hat{r}},
\end{equation}
where now only the radial term is selected, because we are interested in the longitudinal electric field. By substitution of (\ref{eqn:equation1}) into (\ref{eqn:electric2}), then equating the radial terms from the right side of (\ref{eqn:electric2}) to the radial component of the total electric field $\mathbf{E} = E_r \mathbf{\hat{r}} + E_{\theta} \mathbf{\hat{\theta}} + E_{\phi} \mathbf{\hat{\phi}}$, we can find 
\begin{equation}
    E_r = -j \omega A_r \frac{e^{-j(kr-\omega t)}}{r} (1-\frac{k^2 c^2}{\omega^2}) = 0.
\end{equation}
Thus, we can wrongly conclude that the long-range longitudinal electric field is zero. However, this is due to the fact that  (\ref{eqn:electric2}) is valid only in the Lorenz gauge, while (\ref{eqn:electric}) is more general expression valid in any gauge. 

\subsection{Derivation for Arbitrary Gauge}

Let us now use (\ref{eqn:electric}) instead of (\ref{eqn:electric2}). For that, gradient of the scalar potential (\ref{eqn:general_potential}) can be found
\begin{equation}\label{eqn:identity1}
    \nabla V = -j k V \frac{e^{-j(kr-\omega t)}}{r} \mathbf{\hat{r}},
\end{equation}
where again we left only the most significant term along the radial direction. Time differential of the radial component of the vector potential  is
\begin{equation}\label{eqn:identity5}
    \frac{\partial{A_r}}{\partial{t}} \mathbf{\hat{r}} = j \omega A_r \frac{e^{-j(kr-\omega t)}}{r} \mathbf{\hat{r}}.
\end{equation}
If (\ref{eqn:identity1}) and (\ref{eqn:identity5}) are substituted into (\ref{eqn:electric}), then we get for radial component
\begin{equation}\label{eqn:longitudinal_electric}
    E_r = j k (V - A_r c) \frac{e^{-j(kr-\omega t)}}{r} \neq 0.
\end{equation}
It can be observed that in the general case longitudinal electric field is nonzero. It becomes null only if $k V = A_r \omega$ or $V = A_r c$. Let us demonstrate that this identity is valid only in the Lorenz gauge.
Time differential of the scalar potential is
\begin{equation}\label{eqn:identity3}
    \frac{\partial{V}}{\partial{t}} = j V \omega \frac{e^{-j(kr-\omega t)}}{r}.
\end{equation}
If (\ref{eqn:identity3}) and (\ref{eqn:identity2}) are put into (\ref{eqn:lorenz_gauge}), then we get
\begin{equation}\label{eqn:identity4}
    \frac{j V \omega}{c^2} \frac{e^{-j(kr-\omega t)}}{r} = A_r jk \frac{e^{-j(kr-\omega t)}}{r} \Rightarrow \frac{V \omega}{c^2} - A_r k = 0.
\end{equation}
After simple algebraic manipulations, we get from the last equation $V  - A_r c = 0$. Thus, it is clear that under Lorenz gauge condition (\ref{eqn:identity4}), longitudinal electric field (\ref{eqn:longitudinal_electric}) vanishes. However, in the general case or in any other gauge, where $V - A_r c \neq 0$, $E_r$ is nonzero. 

\begin{figure}
	\centering
	\includegraphics[width = 0.9\columnwidth]{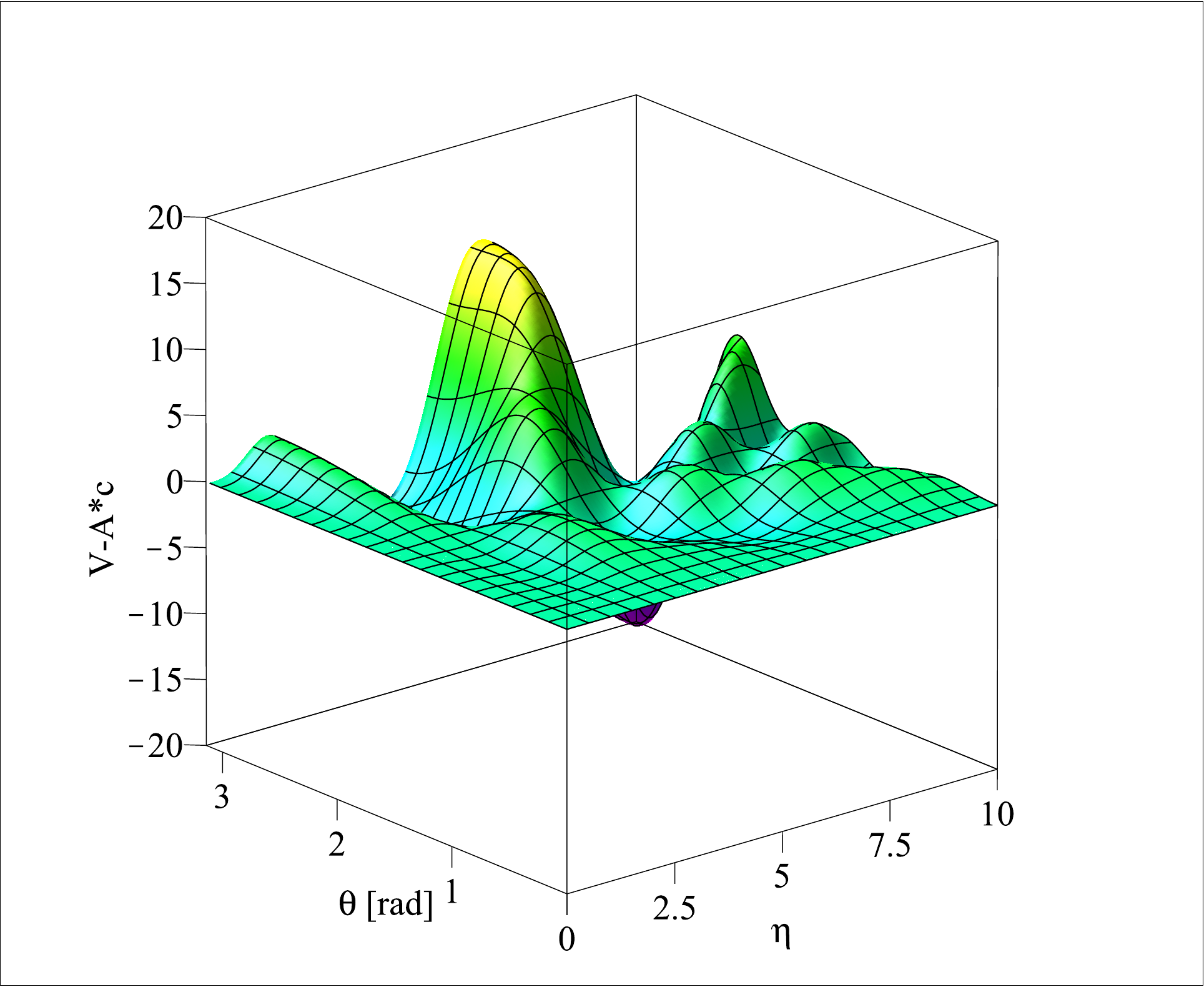}
	\caption{Visualization of $V-A_r c$ as a function of $\theta$ and $\eta = \frac{\omega d}{2c}$. This plot is generated for $\omega(t-\frac{r}{c}) = \frac{\pi}{2}$.}
	\label{fig:lorenz}
\end{figure}

Let us give two examples. For classical electric dipole with charge separation distance $d$ and wavelength $\lambda$, under the assumptions $d \ll \lambda \ll r$, the solutions for emitted scalar potential and radial component of the vector potential are \cite{DavidGriffiths} [Sec. 11.1]
\begin{subequations}\label{eqn:s_and_v}
\begin{equation}
    V = -\frac{q_0}{4\pi \epsilon_0} \frac{\omega d}{r c} \sin{(\omega(t-\frac{r}{c}))} \cos{\theta},
\end{equation}
\begin{equation}
    A_r=-\frac{q_0}{4\pi \epsilon_0} \frac{\omega d}{r c^2} \sin{(\omega(t-\frac{r}{c}))} \cos{\theta}.
\end{equation}
\end{subequations}
Thus it is clear that for the classical dipole $V-A_r c=0$ and as expected $E_r=0$. However, for the electric dipole under the assumptions $\lambda \ll d \ll r$ and with the current wave flowing in positive $\mathbf{\hat{z}}$ direction, the following solutions are obtained \cite{zhakatayev2018longrange}
\begin{subequations}\label{eqn:s_and_v2}
\begin{align}
    & V = -\frac{q_0}{4\pi\epsilon_0} \frac{1}{r} \Bigg( \cos(\omega(t-\frac{r}{c})) \nonumber \\
    & \Big( \cos(\frac{\omega d}{2c}(2-\cos{\theta})) - \cos(\frac{\omega d}{2c}\cos{\theta})  +  \nonumber \\
	& \frac{1}{1+\cos{\theta}} (\cos(\frac{\omega d}{2c})-\cos(\frac{\omega d}{2c}(2+\cos{\theta}))) + \nonumber  \\ & \frac{1}{1-\cos{\theta}}(\cos(\frac{\omega d}{2c}\cos{\theta})-\cos(\frac{\omega d}{2c})) \Big) + \nonumber  \\
	& \sin(\omega(t-\frac{r}{c})) \Big( \sin(\frac{\omega d}{2c}\cos{\theta}) - \sin(\frac{\omega d}{2c}(2-\cos{\theta})) + \nonumber  \\
	& \frac{1}{1+\cos{\theta}} (\sin(\frac{\omega d}{2c}(2+\cos{\theta}))-\sin(\frac{\omega d}{2c})) + \nonumber  \\ 
	& \frac{1}{1-\cos{\theta}} (\sin(\frac{\omega d}{2c})-\sin(\frac{\omega d}{2c}\cos{\theta})) \Big) \Bigg),
\end{align}
\begin{align}
    & A_r=-\frac{q_0}{4\pi \epsilon_0} \frac{1}{rc} (\cos{\theta}) \cdot 	\Bigg( \cos(\omega(t-\frac{r}{c})) \nonumber \\
	& \Big( \frac{1}{1+\cos{\theta}} (\cos(\frac{\omega d}{2c})-\cos(\frac{\omega d}{2c}(2+\cos{\theta}))) + \nonumber \\ & \frac{1}{1-\cos{\theta}}(\cos(\frac{\omega d}{2c}\cos{\theta})-\cos(\frac{\omega d}{2c}))\Big) + \nonumber  \\
	& \sin(\omega(t-\frac{r}{c})) \Big( \frac{1}{1+\cos{\theta}} (\sin(\frac{\omega d}{2c}(2+\cos{\theta}))-\sin(\frac{\omega d}{2c})) +  \nonumber  \\
	& \frac{1}{1-\cos{\theta}} (\sin(\frac{\omega d}{2c})-\sin(\frac{\omega d}{2c}\cos{\theta})) \Big) \Bigg).
\end{align}
\end{subequations}
As a result, 
\begin{equation}\label{eqn:gauge_function}
\begin{aligned}
    & V - A_r c = -\frac{q_0}{4\pi\epsilon_0} \frac{1}{r} \Bigg( \cos(\omega(t-\frac{r}{c}))  \\
    & \Big( \cos(\frac{\omega d}{2c}(2-\cos{\theta})) - \cos(\frac{\omega d}{2c})  +   \\
	& \frac{1-\cos{\theta}}{1+\cos{\theta}} (\cos(\frac{\omega d}{2c})-\cos(\frac{\omega d}{2c}(2+\cos{\theta}))) \Big) +   \\
	& \sin(\omega(t-\frac{r}{c})) \Big( \sin(\frac{\omega d}{2c}) - \sin(\frac{\omega d}{2c}(2-\cos{\theta})) +   \\
	& \frac{1-\cos{\theta}}{1+\cos{\theta}} (\sin(\frac{\omega d}{2c}(2+\cos{\theta}))-\sin(\frac{\omega d}{2c}))  \Big) \Bigg) \neq 0,
\end{aligned}
\end{equation}
Therefore, for electric dipole system where charge separation distance is non-negligible, $V - A_r c \neq 0$ and as a consequence $E_r \neq 0$. Indeed, it can be verified rather easily that if (\ref{eqn:s_and_v2}) are utilized in (\ref{eqn:electric}), then we get non-zero longitudinal electric wave, but if (\ref{eqn:s_and_v2}) are used in (\ref{eqn:electric2}), then due to exact cancellation of all terms, $E_r=0$. Plot of (\ref{eqn:gauge_function}) as a function of $\theta$ and $\frac{\omega d}{2c}$ and for the time when $\omega(t-\frac{r}{c}) = \frac{\pi}{2}$ is shown in Fig.~\ref{fig:lorenz}. This plot looks similar if time is selected such that $\omega(t-\frac{r}{c}) = 0$ or for any other time value. It can be observed that apart of two regions $\frac{\omega d}{2c} \approx 0$ and $\theta \approx 0$, $V-A_r c$ is substantially different than null. We note that applying the infinitesimal electric dipole assumptions ($d \ll \lambda \ll r$) or Lorenz gauge condition lead to $V - A_r c \approx 0$, i.e. lead to disappearance of $E_r$. Similar to our result, it was shown in \cite{rousseaux2003physical} that the identity $V=c A_x$, where $A_x$ is the component of the vector potential along the direction of propagation of plane wave, is responsible for disappearance of the longitudinal electric field. The author also writes that the Fourier transform, when applied to the Lorenz gauge condition, results in the identity $V=c A_x$.

Finally, we can also show that the far-field transverse electric field is caused primarily by the vector potential, while the scalar potential does not contribute to it. The component of the gradient of the scalar potential in the spherical coordinate system along transverse direction is 
\begin{equation}
    \nabla V = \frac{\partial{V}}{\partial{\theta}} \frac{e^{-j(kr-\omega t)}}{r^2} \mathbf{\hat{\theta}},
\end{equation}
where it can be observed that this expression decreases faster than $\frac{1}{r}$. As a result, the far-field transverse electric field arises solely due to contribution of the vector potential 
\begin{equation}
    E_{\theta} \mathbf{\hat{\theta}} = -\frac{\partial{A_{\theta}}}{\partial{t}} \mathbf{\hat{\theta}} = j \omega A_{\theta} \frac{e^{-j(kr-\omega t)}}{r} \mathbf{\hat{\theta}}.
\end{equation}
We can summarize by stating that the far-field transverse electric field is caused by current sources only (vector potential), while the long-range longitudinal electric field is caused by contribution of current and charge sources (vector and scalar potentials), respectively. Similar conclusion applies for $\mathbf{\hat{\phi}}$ direction. For example, for small circular current loop antenna, transverse electric field is along $\mathbf{\hat{\phi}}$, not $\mathbf{\hat{\theta}}$.

\section{Discussion}
\label{sec:discussion}

Even though the presented derivations are simple, their implications are far-reaching. Three main points can be highlighted with respect to obtained results.
\begin{itemize}
    \item Firstly, in many textbooks it is claimed that there is so-called ``gauge freedom'' or gauge invariance \cite{DavidGriffiths,WolfgangPanofsky,smythe1989static,john1999classical,franklin2005classical}. In other words, we can switch from one gauge into another and that would lead to the same electric and magnetic fields. Our derivations challenge this viewpoint. If long-range longitudinal electric field exist in all gauges except the Lorenz gauge, then gauge freedom is no longer valid. It means that if derivations are performed in the Lorenz gauge, then the obtained results would be without longitudinal electric field. However, if derivations are done in the Coulomb or Weyl gauges, then longitudinal electric fields would appear.
    \item The second point is the consequence of the first. Often it is claimed that the vector and scalar potentials are not real physical quantities, but are utilized as mathematical tools \cite{franklin2005classical}. However, there are recent theoretical studies which demonstrate that the vector and scalar potentials carry more information than the electric and magnetic fields \cite{reiss2012modified,reiss2017physical}. Also there are papers where the vector and scalar potential waves were experimentally detected \cite{monstein2002,zimmerman2011macroscopic,zimmerman2013reception,daibo2015vector,daibo2016vector}. All these studies imply that the vector and scalar potentials are real physical fields. That would also explain why ``gauge freedom'' is not valid. The following example can be given with respect to confusion coming from the vector and scalar potentials and their gauge freedom. In classical mechanics, acceleration of any given object is absolute for all inertial reference frames, while its velocity is relative. However, the relative nature of velocity does not imply that it is not real physical quantity. Velocity has different magnitudes in different inertial frames of reference, but at each inertial frame it is real physical quantity. Radius vector of an object is even more relative quantity than velocity, because radius-vector not only depends on the inertial frame of reference, but for a given inertial frame, it also depends on the origin of the frame. If for a given inertial frame, its origin is changed, then the position vector of an object also changes. However, again relative nature of position vector does not imply that it is just mathematical tool used to solve mechanics problems. If logic of gauge freedom is applied in classical mechanics, then we would conclude that acceleration is real physical quantity, because it manifests itself in Newton's laws, while velocity and position are simple mathematical tools. The difference between the vector potential in electrodynamics and velocity, position vectors in mechanics is that the latter quantities can be measured, while there are yet no tools and methods which would permit measurement of the  former physical quantity. 
    \item Thirdly, contrary to established opinion, the obtained results show that far-field longitudinal electric wave might be real. By using the Lorenz gauge condition unintentionally, in some textbooks authors demonstrate that longitudinal electric waves are negligible. For instance, Lorenz gauge condition was utilized to derive radiation of electric dipole in \cite{WolfgangPanofsky} [Sec. 14-3], \cite{ohanian1988classical} [Sec. 14.1], \cite{john1960foundations} [Sec. 16-8], \cite{balanis2005antenna} [Sec. 3.6], \cite{king1969antennas} [Sec. 1.15], \cite{stutzman2013antenna} [Sec. 2.2]. In light of the presented derivations, their conclusions might need to be revised. One possible reason which explains why (\ref{eqn:electric2}) is more frequently used in textbooks than (\ref{eqn:electric}) is that evaluation of the vector potential is sufficient in the former, while both scalar and vector potentials are necessary in the latter case. As a result, when deriving radiation fields of electric dipoles, magnetic dipoles or antennas, it might be more tempting to use (\ref{eqn:electric2}). However, shorter path might mislead and give only part of the answer, not the full answer. 
\end{itemize}





\bibliography{apssamp}

\end{document}